\documentclass[3p,times,twocolumn]{elsarticle}

\usepackage{amssymb}

\usepackage[figuresright]{rotating}

\begin{document}

\begin{frontmatter}




\title{GAMMA-LIGHT: High-Energy Astrophysics above 10 MeV}


\author[label1]{Aldo Morselli}
\author[label2] {Andrea Argan} 
\author[label3,label4] {Guido Barbiellini} 
\author[label3] {Walter Bonvicini} 
\author[label5]  {Andrea Bulgarelli}
\author[label6]  {Martina Cardillo}
\author[label7]  {Andrew Chen}
\author[label8]  {Paolo Coppi}
\author[label9]  {Anna Maria Di Giorgio}
\author[label9]  {Immacolata Donnarumma}
\author[label9]  {Ettore Del Monte}
\author[label5]  {Valentina Fioretti}
\author[label10]  {Marcello Galli}
\author[label9]  {Manuela Giusti}
\author[label11]  {Attilio Ferrari}
\author[label5]  {Fabio Fuschino}
\author[label12]  {Paolo Giommi }
\author[label13]  {Andrea Giuliani}
\author[label5]  {Claudio Labanti}
\author[label12]  {Paolo Lipari}
\author[label3,label4]  {Francesco Longo}
\author[label9]  {Martino Marisaldi}
\author[label9]  {Sergio Molinari}
\author[label13]  {Carlos Mu\~noz}
\author[label14]  {Torsten Neubert}
\author[label15]  {Piotr Orlea\'nski}
\author[label16]  {Josep M. Paredes}
\author[label19]  {M. \'Angeles P\'erez-Garc\'ia }
\author[label9]  {Giovanni Piano}
\author[label1,label6] {Piergiorgio Picozza}
\author[label20]  {Maura Pilia}
\author[label18]  {Carlotta Pittori }
\author[labe19]  {Gianluca Pucella}
\author[label9]  {Sabina Sabatini}
\author[label9,label6,label1]  {Edoardo Striani}
\author[label99,label6,label1]  {Marco Tavani}
\author[label21]  {Alessio Trois}
\author[label3]  {Andrea Vacchi}
\author[label9]  {Stefano Vercellone}
\author[label18]  {Francesco Verrecchia}
\author[label9]  {Valerio Vittorini}
\author[label9]  {Andrzej Zdziarski}

 \address[label1]{INFN Roma Tor Vergata }
\address[label2]{INAF Roma, Italy } 
\address[label3]{INFN Trieste, Italy}
\address[label4]{Univ. of Trieste, Italy }
\address[label5]{INAF-IASF Bologna, Italy}
\address[label6]{Univ. di Roma "Tor Vergata"}
\address[label7]{INAF-IASF Milano, Italy}
\address[label8]{Yale Univ., USA}
\address[label9]{INAF-IAPS Roma, Italy}
\address[label10]{ENEA Bologna, Italy}
\address[label11]{Univ. Torino, Italy}
\address[label12]{ASDC Frascati, Italy}
\address[label13]{INAF-IASF Milano, Italy}
\address[label11]{Univ. Torino, Italy}
\address[label12]{INFN and Univ. di Roma "La Sapienza"}
\address[label13]{Universidad Autonoma de Madrid and IFT-UAM/CSIC, Spain}
\address[label14]{DTU Space, Denmark}
\address[label15]{SRC PAS, Poland}
\address[label16]{Univ. Barcelona, Spain}
\address[label17]{University of Salamanca and IUFFyM, Salamanca, Spain}
\address[label18]{INAF OAR and ASDC}
\address[label19]{ENEA Frascati, Italy}
\address[label20]{ASTRON, The Netherlands}
\address[label21]{INAF OAC, Cagliari, Italy}
\address[label22]{INAF-IASF Palermo, Italy}
\address[label23]{NCAC, Poland)}

\begin{abstract}
The energy range between 10 and 50 MeV is an experimentally very difficult range and remained uncovered since the time of COMPTEL.  Here we propose a possible mission to cover this energy range.
\end{abstract}


\end{frontmatter}


\section{Introduction}
\label{}

High-energy phenomena in the cosmos, and in particular processes leading to the emission of gamma- rays in the energy range 10 MeV - 100 GeV, play a very special role in the understanding of our Universe. This energy range is indeed associated with non-thermal phenomena and challenging particle acceleration processes. The Universe can be thought as a context where fundamental physics, relativistic processes, strong gravity regimes, and plasma instabilities can be explored in a way that is not possible to reproduce in our laboratories. High-energy astrophysics and atmospheric plasma physics are indeed not esoteric subjects, but are strongly linked with our daily life. Understanding cosmic high-energy processes has a large impact on our theories and laboratories applications. The technology involved in detecting gamma-rays is challenging and drives our ability to develop improved instruments for a large variety of applications.
GAMMA-LIGHT is a Small Mission which aims at an unprecedented advance of our knowledge in many sectors of astrophysical and Earth studies research. The Mission will open a new observational window in the low-energy gamma-ray range 10-50 MeV, and is configured to make substantial advances compared with the previous and current gamma-ray experiments (AGILE \cite{Agile} and Fermi {\cite{Fermi}). The improvement is based on an exquisite angular resolution achieved by GAMMA-LIGHT using state-of-the-art Silicon technology with innovative data acquisition.
Despite the recent important results and progress, AGILE and Fermi
are leaving crucial unresolved issues in the energy window 10-100 MeV.
 GAMMA-LIGHT will address all astrophysics issues left open by the current generation of instruments. In particular, the breakthrough angular resolution in the energy range 100 MeV - 1 GeV is crucial to resolve patchy and complex features of diffuse sources in the Galaxy as well as increasing the point source sensitivity. This proposal addresses scientific topics of great interest to the community, with particular emphasis on multifrequency correlation studies involving radio, optical, IR, X-ray, soft gamma-ray and TeV emission. At the end of this decade several new observatories will be operational including LOFAR, SKA, ALMA, HAWK, CTA. GAMMA-LIGHT will "fill the vacuum" in the 10 MeV-10 GeV band, and will provide invaluable data for the understanding of cosmic and terrestrial high-energy sources.

\section{Astrophysics Objectives of GAMMA-LIGHT}
\label{}

Many crucial scientific issues are left unsolved by the current generation of gamma-ray instruments (AGILE, Fermi). They constitute the main GAMMA-LIGHT scientific objectives :

1. more sensitive search of Dark Matter gamma-ray signatures in the Galaxy and in particular in the Galactic Center region;

2. completely resolving the Galactic Center region in gamma-rays: Sgr A*, GeV and TeV sources, nebulae, compact sources, SNRs;

3. resolving the diffuse emission in the Galactic plane in relation with cosmic-ray propagation, star forming regions in the Galactic plane; extending the cosmic-ray propagation and emission properties of the "Fermi bubbles" to the lowest energies below 100 MeV;

4. resolving spatially and spectrally SNRs and addressing the origin and propagation of cosmic- rays with unprecedented accuracy;

5. polarization studies of gamma-ray sources;

6. detection of soft gamma-ray pulsars in the range 10-100 MeV, and pulsar wind nebulae studies;

7. detection of compact objects, microquasars, relativistic jets in the range 10 MeV - 1 GeV
resolving the issue of hadronic vs. leptonic jets for a variety of sources (e.g., Cyg X-3);

8. detection and localization of transients and exotic sources with much improved sensitivity;
detection of Crab Nebula gamma-ray flares with excellent sensitivity down to 10 MeV;

9. blazar studies down to 10 MeV, excellent positioning resolving source confusion;

10. GRB excellent capability in the range 10 MeV - 5 GeV; sub-millisecond timing capability in the
range 0.3-100 MeV.

\begin{figure}
\centering
\includegraphics[width=0.5\textwidth]{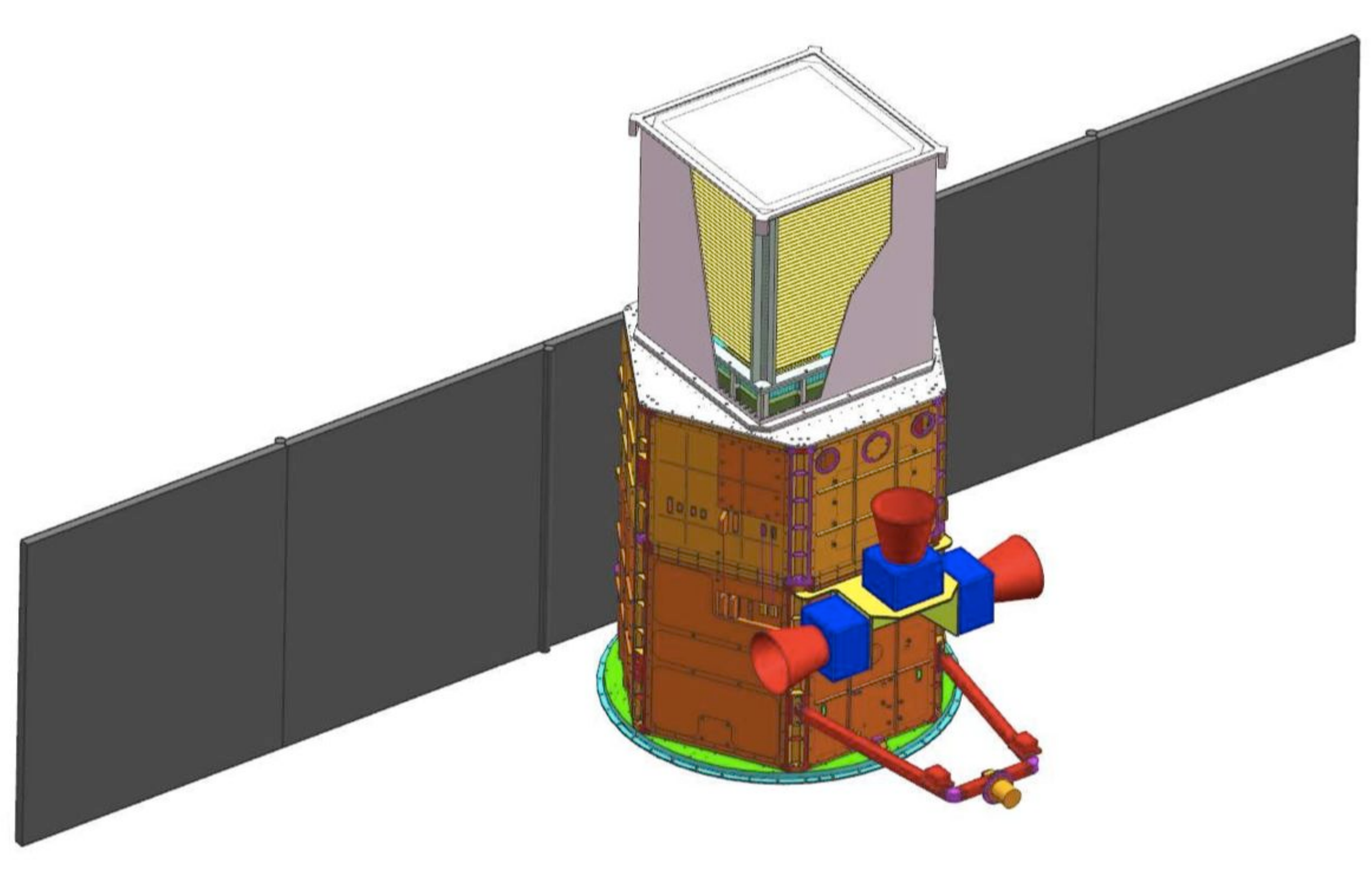}
\caption{GAMMA-LIGHT scheme. On the top there is the  scientific instrument showing the external AC system, the Silicon Tracker, and the Calorimeter.\label{fig:instr}}
\end{figure}

\section{The instrument}

A scheme of GAMMA-LIGHT can be seen in figure~\ref{fig:instr}.
The gamma-ray Tracker is the heart of the  payload and it is made of 40 planes of silicon strip detectors organized in 41 trays without  tungsten converter. Each tray is configured as follows: two layers of 25 Silicon tiles each (except trays 1 and 41 which have a single layer) organized in 5 ladders composed of 5 tiles bonded together. 
Each tray is made of a 1 cm core of aluminum honeycomb covered on both sides by a 0.5 mm thick Carbon fiber layer. The resulting tray height is 1.1 cm and the total Tracker height is slightly above 50 cm. The Tracker is a compact, low-power 153.600 channel detector with self-triggering capability, fast timing possibility and full analog readout. The active element is a single-sided, AC-coupled, 410~$ \mu$m thick, 9.5x9.5 cm$^2$ Silicon strip detector with microstrips of 121 $\mu$m  pitch, alternate strip readout pitch of 242~$\mu$m with one floating strip, and polysilicon resistors for the bias.

The Calorimeter (CAL) is made of CsI elements each of dimensions: height = 4.5 cm, square size of 1 cm. Total lateral length of 50 cm (48x48 pixels).

The Anticoincidence (AC) system is made of a top plastic scintillator  layer of thickness of 0.6 cm and lateral panels surrounding all four sides (3 panels for each side). All panels are coupled to PMTs by scintillating optical fibers.

\begin{figure}
\centering
\includegraphics[width=0.49\textwidth]{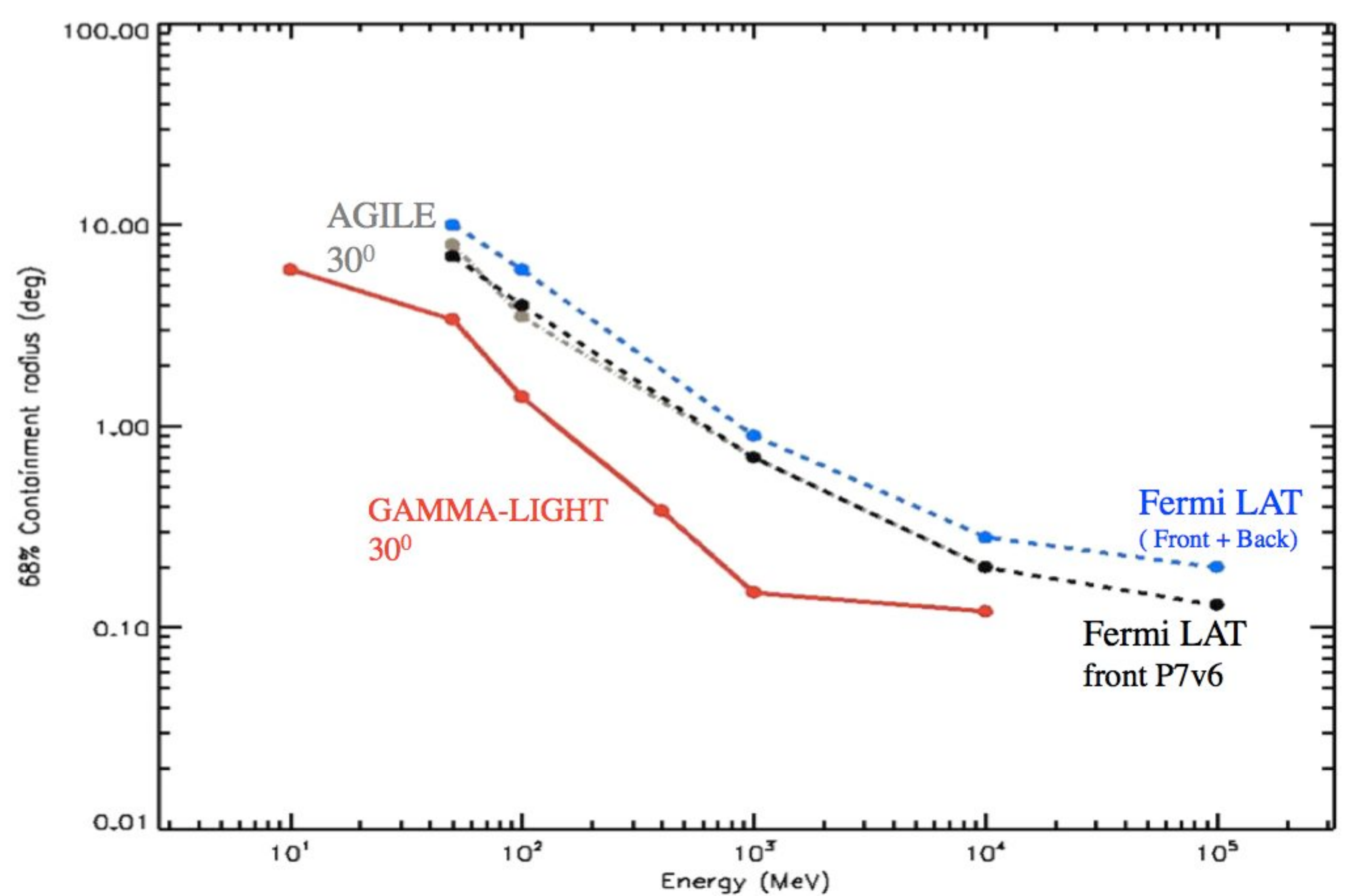}
\caption{Point Spread Function (PSF, 68$\%$ containment radius) of the GAMMA-LIGHT gamma-ray (GRID) imager (in red color) obtained by extensive GEANT-4 simulations which assume an incidence angle of 30$^{\circ}$, Silicon strip analog readout, and Kalman filter analysis of particle tracks. For comparison, we show the Fermi-LAT Pass7V6 PSF (total LAT: blue curve; front-LAT: black color) and the AGILE PSF (in gray color).\label{fig:ang}}
\end{figure}

\begin{figure}[ht]
\centering
\includegraphics[width=0.49\textwidth]{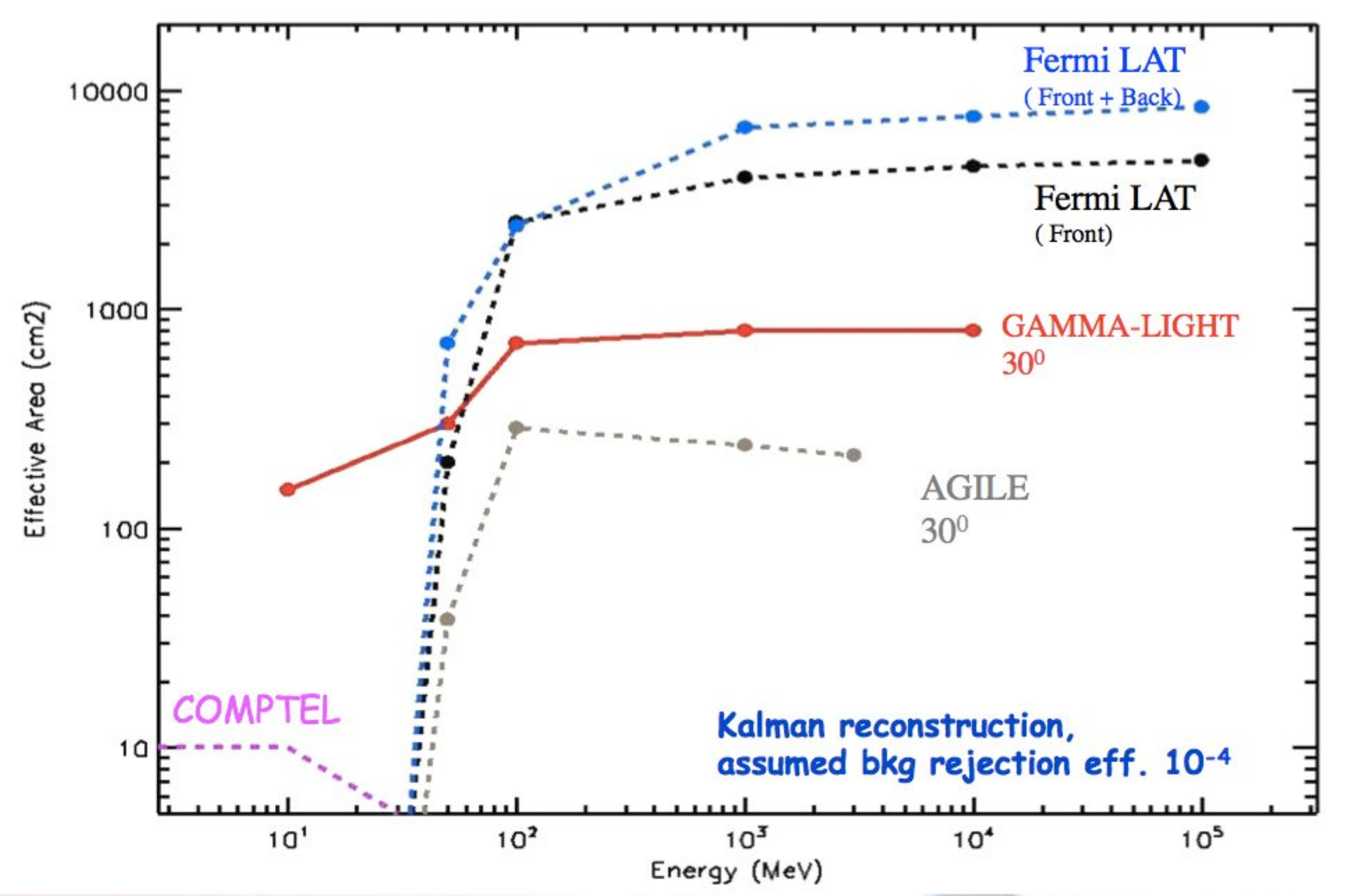}
\caption{Effective  area  for  the  GAMMA-LIGHT  GRID  at  30  degree
off-axis  (in  red  color).  For  comparison,  we  also  show  the
effective  areas  of  AGILE  at  30  degree  off-axis  (in  gray
color),  Fermi-LAT-front  Pass7 V6  at  normal  incidence  (total:
blue  color;  front-LAT:  black  color),  and  COMPTEL's  (in
purple).  Trigger  logic  efficiency  and  background  rejection
have  been  taken  into  account. 
 \label{fig:aeff}}
\centering
\includegraphics[width=0.49\textwidth]{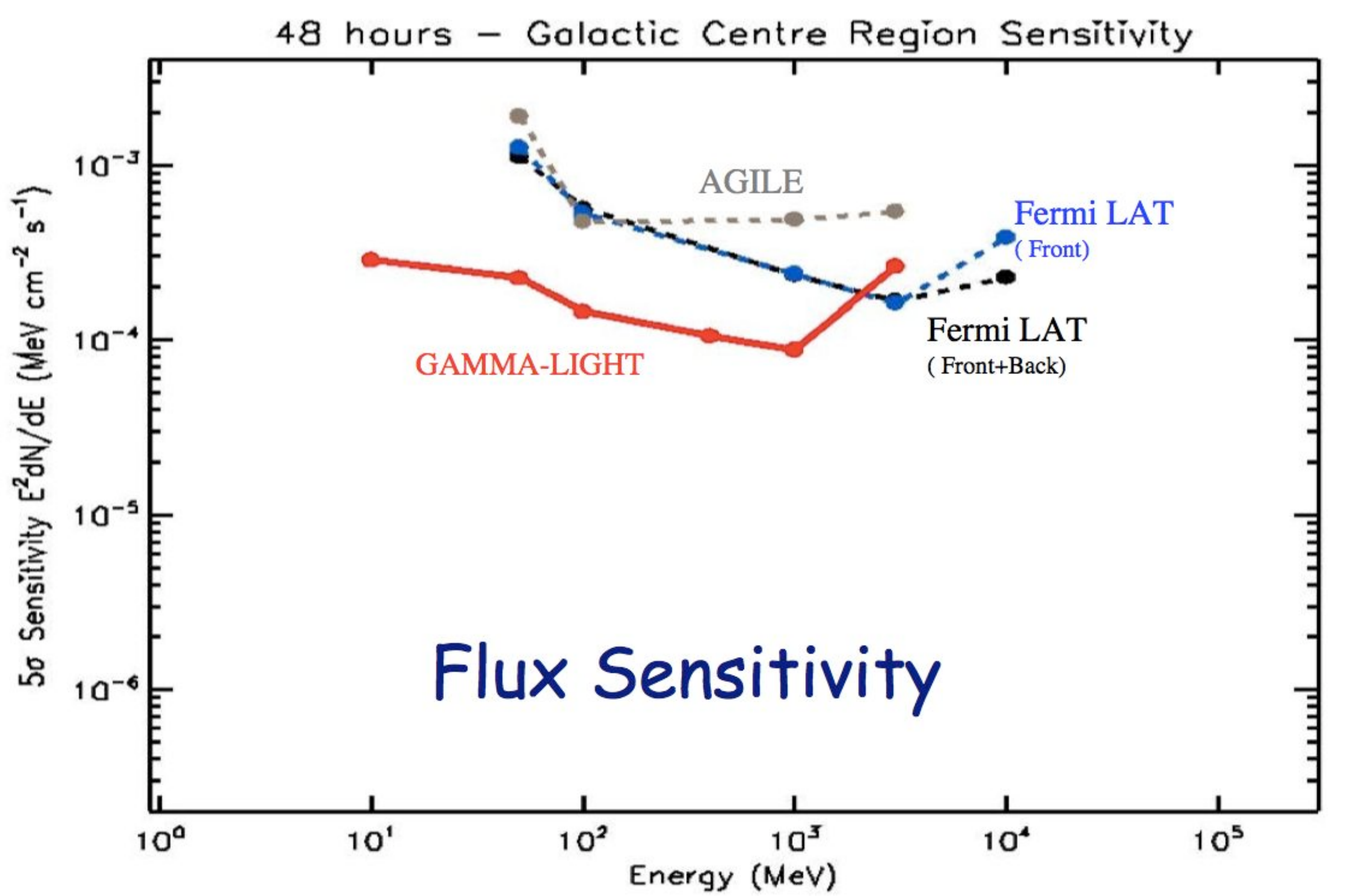}
\caption{Point source (5-sigma) sensitivity for 48 hr (solar time) observation at 30$^{\circ}$ off-axis of the GAMMA-LIGHT GRID imager (in red color). Also shown are the Fermi-LAT Pass7V6 sensitivity (total-LAT: black color; front-LAT: blue color) and AGILE's sensitivity (gray) for the same duration. \label{fig:sens}}
\end{figure}

The Point Spread Function is shown in figure~\ref{fig:ang} , the effective area is shown in figure~\ref{fig:aeff}
and the sensitivity  for 48 hr (solar time) observation is shown in figure~\ref{fig:sens}.  For a 2-day observation, sensitivies up to GeV energies are background dominated. At higher energies, sensitivities are photon limited: here we show the limit sensitivities assuming at least N=5 high-energy photons detected within the 99$\%$ confidence radius. GAMMA-LIGHT is assumed to be pointing in a LEO orbit with an overall exposure efficiency of 0.6 similar to AGILE's (checked with real data). Fermi-LAT is assumed to be in sky-scanning mode with an overall exposure efficiency per single source of 0.16 (as checked with real data). Final data acquisition efficiencies are assumed to be equal to 0.6; they take into account background rejection in a LEO orbit and on-board trigger logic and ground data processing as deduced from Fermi and AGILE.

\begin{figure}
\centering
\includegraphics[width=0.49\textwidth]{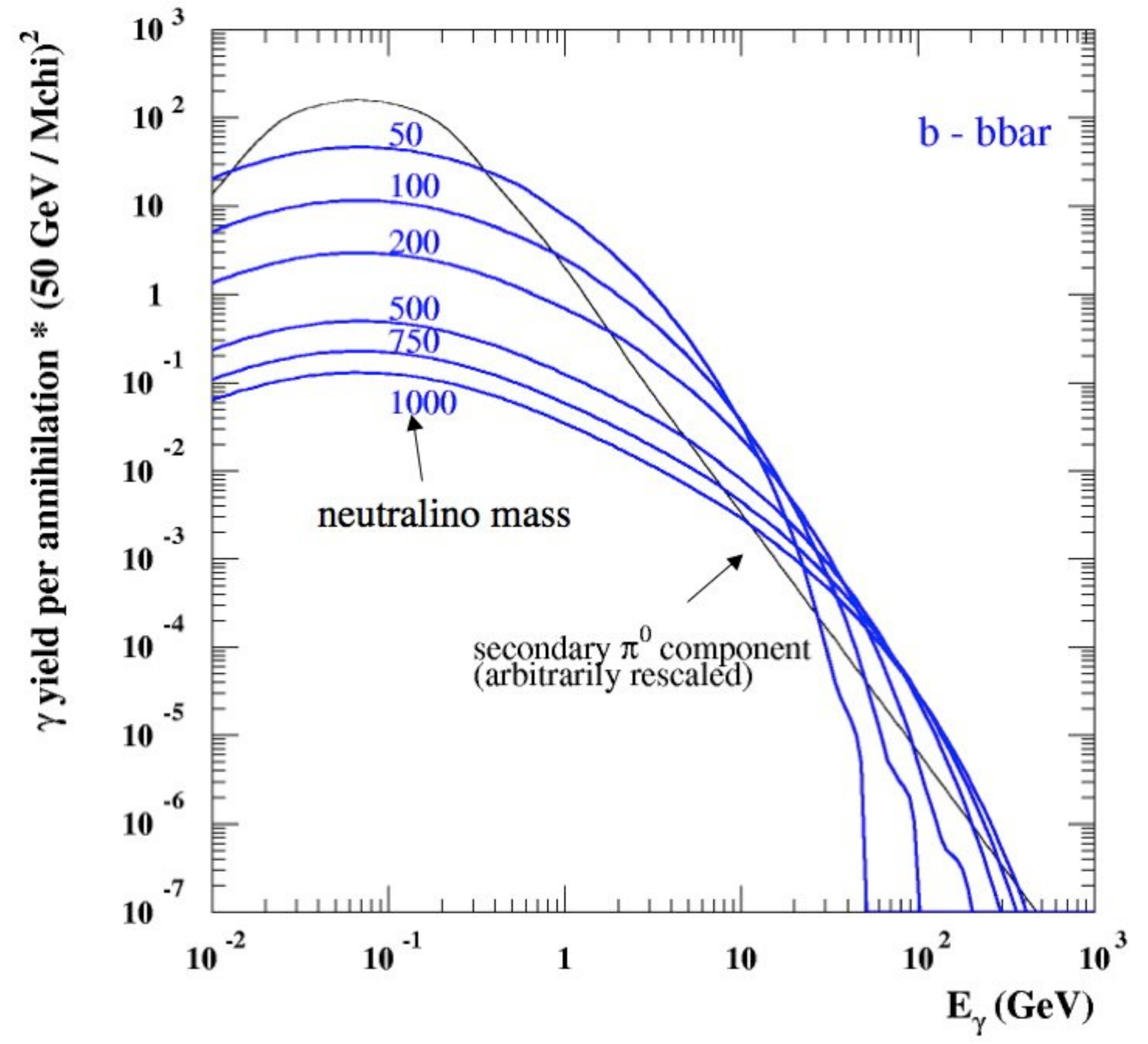}
\caption{Differential $\gamma$-ray energy spectra per annihilation  for a fixed annihilation channel (b bar) and for a few sample values of WIMP masses \cite{dark}. For comparison we also show the emissivity, with an arbitrarily rescaled normalization, from the interaction of primaries with the interstellar medium.  The solid lines are the total yields, while the dashed lines are components not due to $\pi^0$ decays. \label{fig:yield}}
\end{figure}

\section{Dark Matter in our Galaxy}
\label{}

The nature of Dark Matter (DM) is still a mystery. Gamma-ray emission from our Galaxy may reveal the existence of certain types of DM, by means of the production of secondary $\gamma$-rays after the annihilation (or decay) of the DM particle candidates \cite{prelaunch}. 

The importance of GAMMA-LIGHT for Dark Matter searches can be seen in figures~\ref{fig:yield} and \ref{fig:yield2}  where the differential $\gamma$-ray energy spectra per annihilation of  Weakly Interacting Massive  Particle (WIMP) are plotted \cite{dark}. 
As one can see the bulk  of the emission even for high WIMP masses is in the energy range  5 MeV - 100 MeV.

In the Fermi-LAT analysis of the Galactic Center the diffuse $\gamma$-ray backgrounds and discrete sources, as we model them today, can account for the large majority of the 
detected $\gamma$-ray emission from the Galactic Center. Nevertheless a residual  emission is left, not accounted for by the above models  of standard astrophysical phenomena \cite{symp}, \cite{GC_cim},  \cite{GC_Nim}.

So in the  inner region of the Milky Way  a better angular resolution with respect to AGILE  and Fermi   is needed in  the 5 MeV - 100 MeV energy range 
in order to disentangle the possible DM contribution from the  diffuse background and the point sources contribution (see for example \cite{GC_Neur}).

Let us finally remark that decaying DM can produce a detectable line in the Gamma-Light energy range 
\cite{gravitino}. In principle, detectability is expected to be large in the very Galactic Center since hadronic emission models for this region are predicting a fall down about 100 MeV (see Fig. 2 of \cite{figure}).

\begin{figure}
\centering
\includegraphics[width=0.49\textwidth]{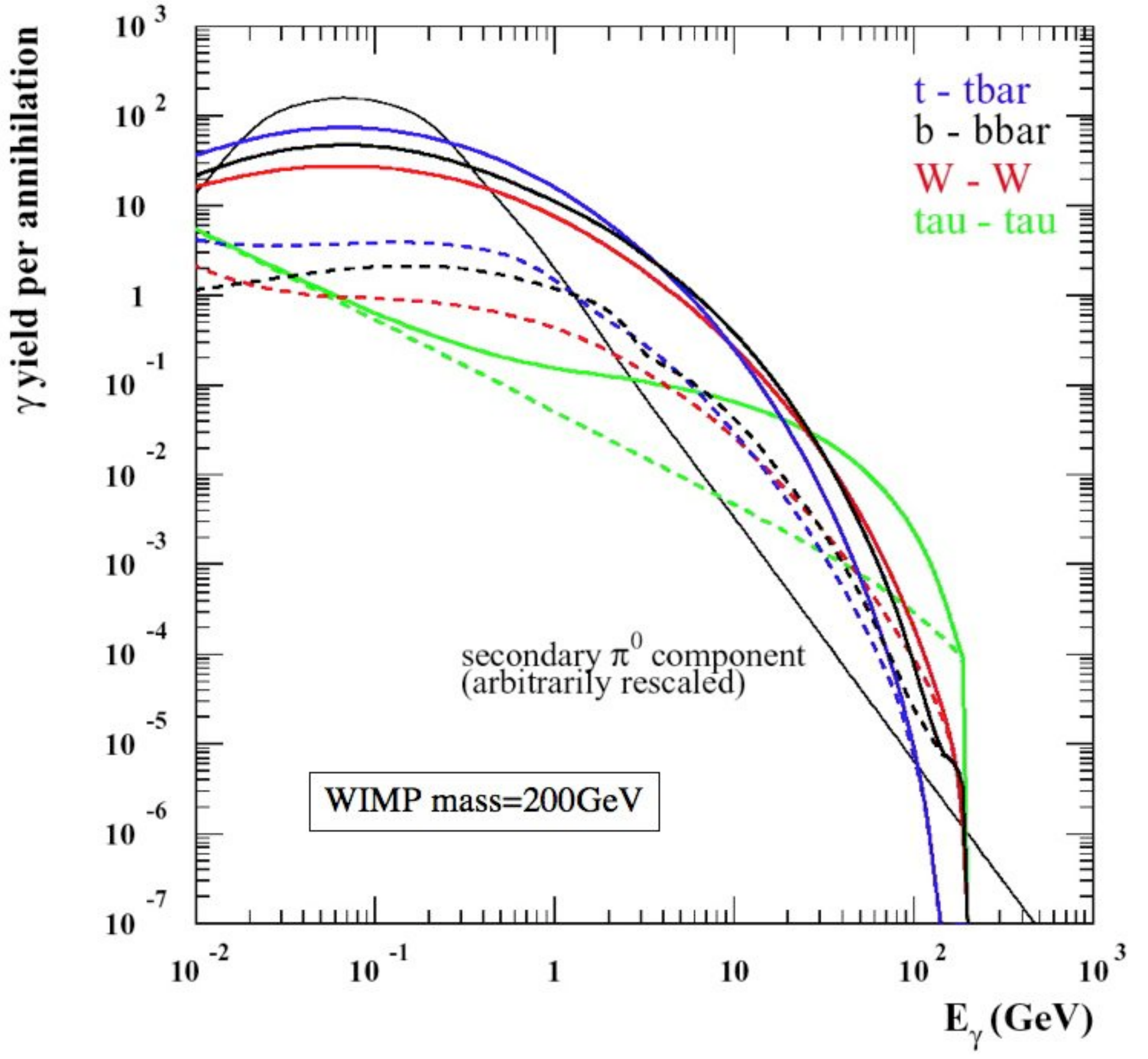}
\caption{Differential energy spectra per annihilation like in figure~\ref{fig:yield} for a few sample annihilation channels and a fixed WIMP mass (200 GeV).\label{fig:yield2}}
\end{figure}

GAMMA-LIGHT can  resolve the Galactic Center (GC) region and similar complex regions of the plane. The GC is indeed one of the most difficult regions to observe in high-energy gamma-rays. Optical emission is heavily obscured by dust, and both the concentration of point sources and the concentration of clouds and diffuse emission enhancements is very high, a fact that complicates both the analysis and source identifications. Towards the GC (and the anti-center) the rotational velocity of the Galaxy is entirely transverse, no longer allowing the distance of the interstellar gas to be determined through radio line shifts. In addition, the column density of atomic hydrogen can be very large. Nevertheless, important progress has been made by AGILE and Fermi. The most surprising discovery has been the observation of large, well-defined bubbles/lobes above and below the Galactic plane extending 
50$^{\circ}$ above and below the GC with a width of about 40¡ degrees in longitude  \cite{Su}. These lobes have a uniform surface brightness with sharp edges, neither limb-brightened nor centrally-brightened, and are nearly coincident with a similar ÒhazeÓ discovered in WMAP data and later confirmed by Planck, suggesting a common origin, most likely inverse Compton emission of high-energy electrons scattering off the microwave photons producing gamma-rays or escaping hadrons. The leptonic or hadronic hypotheses would imply an injection of high-energy particles in the past $\sim$10 Myr. Possible mechanisms include an accretion event onto the central black hole, a nuclear starburst, or the accumulation of events from a precessing jet. Each of these scenarios poses a number of problems. GAMMA-LIGHT will determine the morphology and spectral properties below 1 GeV of the GC region and Fermi-bubbles with unprecedented accuracy, and will contribute to resolving the issue of the nature of this emission.
An example of the difficulty in the analysis of the GC region is the search for the gamma-ray counterpart to the super-massive black hole at the center of the galaxy, Sgr~A*. At TeV energies, HESS has found a strong point source within 10 arcminutes of Sgr~A*. However, source of the TeV emission may be either Sgr A* itself, or a nearby plerion discovered within the central few arcseconds, or a putative "black hole plerion" produced by the wind from Sgr~A*, or the diffuse ²10 pc region surrounding Sgr A*. An analysis of 25 months of Fermi data by  \cite{Chernyakova}  found 4 new sources within a 10$^{\circ}$x10$^{\circ}$ region around Sgr~A* in addition to the 19 already listed in the first Fermi Source Catalog (1FGL). The source coincident with both the HESS source and the position of Sgr~A*, 1FGL J1745.6Ð2900, shows no variability in either GeV or TeV energies, while the GeV spectrum indicates that the emission mechanism must be distinct from that of the TeV emission. Much higher angular resolution is needed to distinguish among the various scenarios, and GAMMA-LIGHT is the ideal instrument to resolve these issues.

\begin{figure}[ht]
\vskip-0.1cm
\centering
\includegraphics[width=0.49\textwidth]{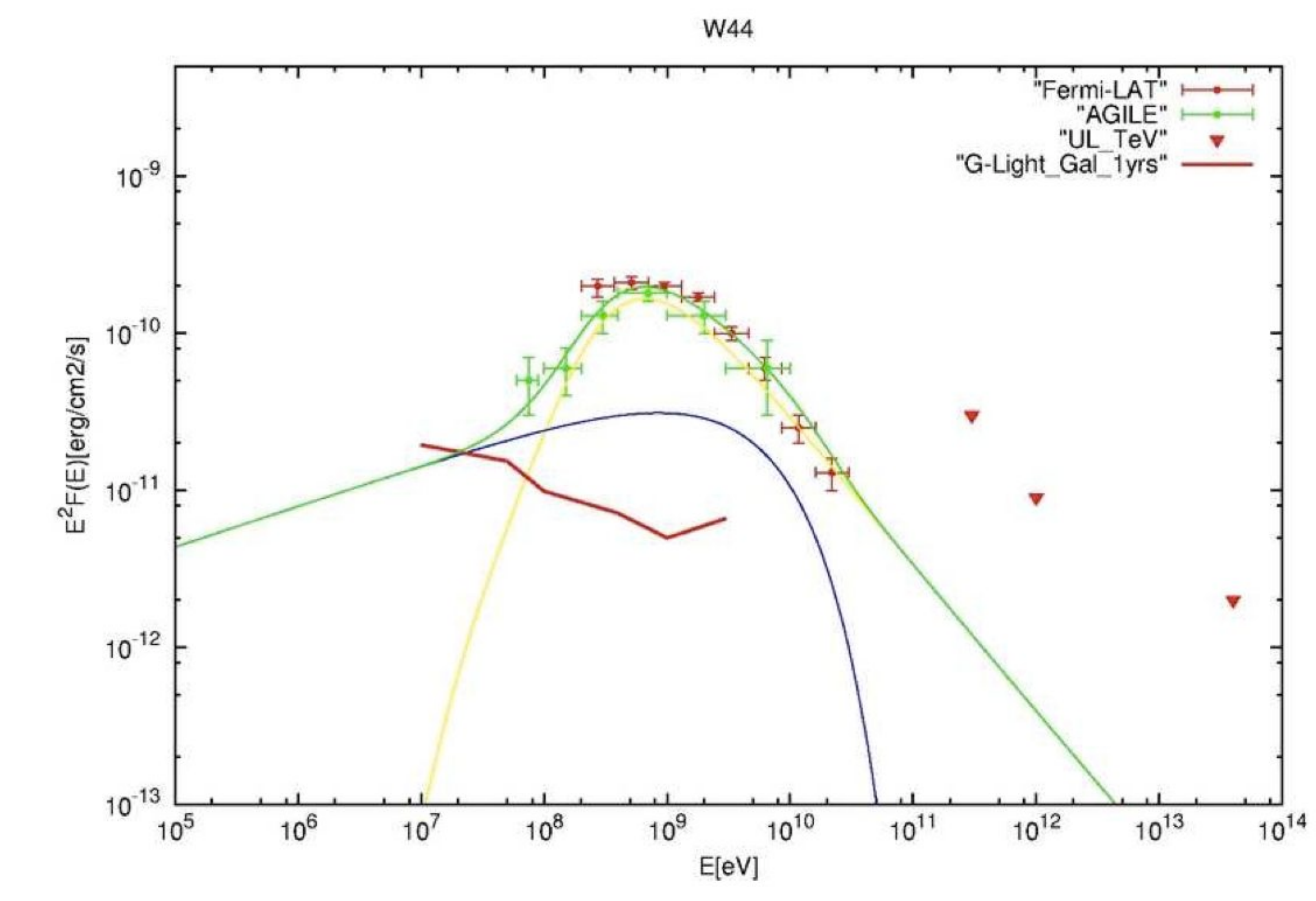}
\caption{The gamma-ray spectrum of one of the most prominent gamma-ray emitting SNRs showing a clear signature of neutral pion decay, W44. The Fermi (green triangles) and AGILE (red diamonds) data are shown. The spectral modeling involves a component due to the neutral pion decay (yellow line) plus a component due to Bremsstrahlung (blue line). The red curve shows the expected GAMMA-LIGHT sensitivity for a 1-year effective time integration.\label{fig:W44}}
\end{figure}

\section{SNRs and Diffuse "Cocoons" in the Galaxy: Origin and Propagation of CRs}

Although a fair number of Supernova Remnants (SNRs) have been detected and recently studied by AGILE and Fermi, a number of outstanding issues regarding the origin of cosmic-rays (CRs) remain to be addressed. Today a few dozen SNRs are known to emit gamma rays: the study of these objects together with their non-thermal properties (e.g., radio and X-ray emission) is the topic of many investigations in CR physics and particle acceleration mechanisms (e.g., \cite{Aharonian} , \cite{Fermi_Science}, \cite{Eta-Carinae_Agile}). However it is often quite difficult to distinguish, for these objects, the different components that contribute to the gamma-ray spectrum (for energies greater than $\sim$10 MeV up to tens of TeV, the only emission processes expected to product emission are the decay of neutral pions produced in p-p scattering, inverse Compton on low energy photons and Bremsstrahlung). The models so far produced to explain the SNR gamma-ray emission show that the knowledge of the spectrum at low energies (below 100 MeV) is crucial for this purpose. In fact, the rapid fall of the gamma-ray spectrum at energies less than 100 MeV is the most significant feature in the spectrum which may allow to discriminate hadronic vs. leptonic emission. Electron Bremsstrahlung can provide an opposite behavior. Until now, this analysis was possible only for very few SNRs (e.g., \cite{W44}, \cite{Uchiyama}). GAMMA-LIGHT will be able to resolve the complex morphology of the SNR gamma- ray emission. It will provide invaluable information for a detailed modelling of CR acceleration and propagation.

 \section{Polarization  Studies} 

 GAMMA-LIGHT  can  greatly  contribute  to  determine  gamma-ray  polarization  for  intense  sources.  The  absence of  high-Z  converters  in  the  gamma-ray  Tracker  makes  possible  the  measurement  of  the  pair  production plane  and  angles  with  good  accuracy.  A  method  to  determine  the  polarization  direction  of  high  energy (50  MeV-30  GeV)  linearly  polarized  gamma-rays  was  analyzed  and  discussed in \cite{Depaola}.  The  polarization  information  is  contained  in  the  azimuthal  distribution  of  the  created  pair.  The  GAMMA-LIGHT
Tracker  will  have  an  excellent  angular  resolution  above a  few  hundreds  of  MeV  to  determine  the  aperture  angle of  positrons  and  electrons.  This is an  exciting possibility  in  the  study  of  intense  gamma-ray  sources.

 \section{Terrestrial Gamma-Ray Flashes}

Terrestrial Gamma-Ray Flashes (TGFs) (see \cite{Dwyer}  for a recent review) are one of the most intriguing phenomena in the geophysical sciences and the manifestation of the highest energy natural particle accelerators on Earth. TGFs are millisecond time-scale bursts of gamma-rays produced above thunderstorms and associated to lightning activity. Although several observations are available and a general picture of this phenomenon based on Bremsstrahlung by relativistic runaway electrons produced in thunderstorms strong electric field is commonly accepted, there are many points which remain obscure. Among the outstanding issues, we mention here the highest achievable energies (which translates into the maximum voltage drop that can be established within thunderclouds) the connection with lightning and cloud microphysics, and the global and local TGF occurrence rate. There are currently three active space instruments capable of TGF detection: AGILE, RHESSI, and Fermi GBM. AGILE in particular have shown that the TGF energy spectrum extends well above 40 MeV (\cite{Marisaldi}  up to 100 MeV  \cite{TGF_Agile} with a spectral shape which is difficult to reconcile with current production models. TGFs are now established as one of the coupling mechanisms between lower and upper atmosphere. Since TGFs appear to be a much more common phenomenon than previously expected \cite{TGF}, it is important to assess the impact of these phenomena on the physical/chemical state of the atmosphere, and on the climate. This is especially true if a large fraction is emitted at high energy, as suggested by the recent AGILE observations, since high-energy particles can have a significant role in aerosol nucleation and ultimately in cloud formation \cite{Kirkby}.
The study of TGFs and energetic radiation from thunderstorms has now entered a golden age. Future observational breakthrough will come with the Atmosphere-Space Interaction Monitor (ASIM), the ESA mission for the study of TGFs and Transient Luminous Events (TLEs), and by the CNES micro-satellite TARANIS, expected to be launched in 2014 and 2015 respectively. However, none of the forthcoming missions have detection capabilities above 40 MeV or imaging capabilities above 10 MeV, where the memory of the electric field orientation at the source region is carried off by gamma-ray photons.

\begin{figure}
\centering
\includegraphics[width=0.49\textwidth]{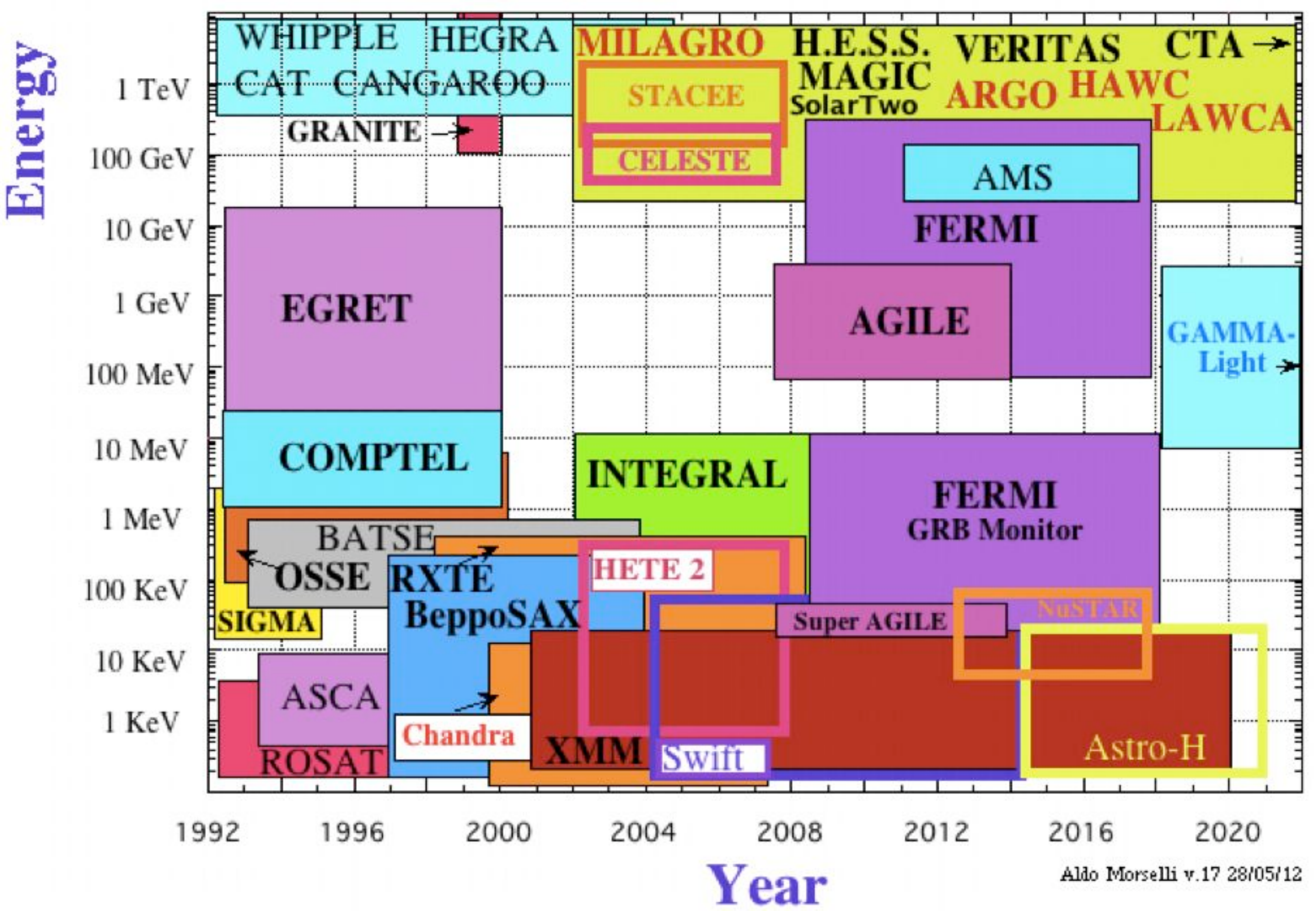}
\caption{Timeline schedule versus the energy range covered by present and future
detectors in X and gamma-ray astrophysics.\label{fig:timel}}
\end{figure}

 \section{Conclusion}
For  X-ray and
gamma-ray experiments  the time
of operation versus  energy range is shown in figure~\ref{fig:timel}.  Note that GAMMA-LIGHT  will cover an interval not covered by any other
experiments. Note also the number  of other experiments at other frequencies that will allow
extensive multifrequency studies.

\end{document}